\documentclass[aps,pre,reprint,superscriptaddress,amsmath,amssymb,floatfix,longbibliography]{revtex4-2}

\usepackage{graphicx}
\usepackage{float} 
\usepackage{amsmath,amssymb}
\usepackage{bm}
\usepackage{booktabs}
\usepackage{microtype}
\usepackage[T1]{fontenc}
\usepackage{times}
\usepackage[colorlinks=true,citecolor=blue,linkcolor=blue,urlcolor=blue]{hyperref}
\usepackage{xcolor}
\usepackage[normalem]{ulem}

\newcommand{\avg}[1]{\langle #1 \rangle}
\newcommand{\Neff}{N_{\mathrm{eff}}}

\begin{document}

\title{Noise-resilient collective order through temporal memory in self-propelled particles}

\author{Mihir Durve}
\email{mihir.durve@iit.it}
\affiliation{Center for Life Nano- \& Neuro-Science, Fondazione Istituto Italiano di Tecnologia (IIT), Rome, Italy}
\author{Sauro Succi}
\affiliation{Center for Life Nano- \& Neuro-Science, Fondazione Istituto Italiano di Tecnologia (IIT), Rome, Italy}
\author{Antonio Celani}
\affiliation{Quantitative Life Sciences section, The Abdus Salam International Center for Theoretical Physics (ICTP), Trieste, Italy}
\affiliation{Department of Oncology, Universit\`a degli Studi di Torino, Torino, Italy}
\affiliation{INFN, Sezione di Torino, Torino, Italy}

\date{\today}

\begin{abstract} 
Temporal memory can stabilise collective motion by filtering noisy local observations, but persistent information can also impede adaptation when the collective state changes. We investigate this competition in Vicsek-type systems using two distinct alignment rules: a finite buffer of past observations and a recursive update of an internal directional state. In both models, increasing effective memory shifts the loss of polar order to stronger noise and can restore ordered motion under conditions for which the corresponding memoryless Vicsek model is disordered. The stationary responses of the two models show an approximate correspondence when parameterised by a kernel-based effective memory, consistent with temporal averaging playing a common role in their ordering behaviour rather than from the details of the memory implementation. Their transient dynamics, however, are not determined by effective memory alone. Following a collective perturbation, recovery is fastest at an intermediate memory depth and becomes slower when outdated observations are retained too strongly. Temporal integration therefore produces a trade-off: deeper memory improves noise resilience, whereas excessive persistence reduces responsiveness. These results identify memory depth as a control parameter that separates the optimisation of stationary order from that of dynamical recovery in noisy active systems. 
\end{abstract}
\maketitle

\section{Introduction}
\label{sec:intro}

The Vicsek model provides a minimal framework for studying collective motion generated by local alignment and noise~\cite{Vicsek1995,Vicsek2012}. As a foundational framework for far-from-equilibrium systems, it accounts for key density fluctuations and advective transport underlying collective behavior in bacterial colonies, migrating cells, active colloids, and self-propelled rods~\cite{Czirok1996,Szabo2006,Peruani2012,Buttinoni2013,Ginelli2010}, thereby placing Vicsek-type dynamics firmly within the broader framework of motile active matter~\cite{Bechinger2016,Gompper2020,Toner1995,Toner1998,Ramaswamy2010,Marchetti2013}.

Collective behaviour is sensitive to how agents acquire and process local information. In Vicsek-type models, the apparent order--disorder transition depends on system size, density inhomogeneities, update rules, and noise prescription~\cite{Gregoire2004,Chate2008,Aldana2007,Baglietto2009}. Restricting the angular interaction range can produce a first-order transition~\cite{Durve2016}, while elementary cognitive rules can generate incidental leaders, nematic patterns, and aggregates~\cite{Barberis2016}.

The standard Vicsek rule uses only the current local heading, whereas real agents may retain and combine earlier observations. Measurements of birds and fish reveal structured neighbourhoods, directional interactions, correlated fluctuations, and nontrivial information transfer~\cite{Ballerini2008,Bialek2012,Cavagna2010,HerbertRead2011,Katz2011}.  Collective dynamics may also become intermittent when individuals balance competing behavioural demands, as observed in sheep herds~\cite{Ginelli2015}. These observations motivate models in which alignment depends on more than an instantaneous estimate of the local direction. 

Several studies have introduced temporal effects into models of collective motion through different mechanisms.  Explicit interaction delays can destabilise a coherently moving swarm and generate oscillatory dynamics~\cite{Forgoston2008}. Inertial descriptions of flocking retain angular momentum, allowing turning information to propagate through an ordered group rather than being absorbed by an instantaneous overdamped response~\cite{Cavagna2015,Attanasi2014}.  Memory of a particle's own past motion can produce collective states that differ from those generated by instantaneous alignment alone~\cite{Nagai2015}.  Delayed and non-reciprocal interactions can also drive condensation and aggregation~\cite{Durve2018},while collective memory has been connected to spatial sorting and persistent group structure~\cite{Couzin2002}. Nonsymmetric interactions provide another route to large collective reorientations, producing coherent swings even in globally ordered systems~\cite{Cavagna2017}. 

Of particular relevance to the present work is the relation between temporal persistence and responsiveness.  Gei{\ss}et al. ~\cite{Geiss2022DelayVicsek} examined delayed orientational interactions in the Vicsek model and found that increasing the delay improves stability against random orientation fluctuations while reducing the ability of the flock to follow a leader.  Chen and Zheng~\cite{ChenZheng2024AdaptiveDelay} investigated a related persistence--responsiveness trade-off and showed that adaptive time delays can improve responsiveness while maintaining persistence.  These results demonstrate that temporal persistence can have competing effects on the stability and response of collectively moving systems. 

Here we examine this interplay for a different form of memory, in which particles integrate noisy local-heading observations rather than respond to a delayed state. We consider two implementations: a finite buffer memory(FBM) model that combines past observations using exponentially decaying weights, and a recursive learning model (RLM) that updates a single internal directional state.  We investigate how memory depth affects stationary polarisation under angular noise and the time required to recover collective order after a prescribed perturbation of the stored directional information.  Unlike studies of response to a leader or an externally imposed change in direction, the recovery measurements here concern the replacement of perturbed internal memory. Comparing the two implementations also allows us to examine whether their stationary and transient behaviour can be related through a common effective memory depth. These questions are relevant to decentralised roboticand multi-agent systems operating with noisy local sensing~\cite{Brambilla2013,Reynolds1987,OlfatiSaber2006,Turgut2008} as well as to droplet motion in multiphase microfluidics \cite{Montessori2019, Sbragaglia2012}. 

The paper is organised as follows. We first introduce the finite-history model, in which noisy local headings are combined using exponentially decaying weights, and define the relevant observables and numerical protocol. We then characterise the finite-system loss of polarisation over several noise amplitudes and particle numbers; the corresponding Binder cumulant and polarisation fluctuations are reported in Appendix~\ref{app:fluctuations}. Next, we examine the effect of memory depth on stationary polarisation at fixed high noise and measure recovery after a prescribed disturbance. We subsequently introduce the RLM, in which a single internal vector is updated recursively without storing individual headings. Finally, we compare the two rules through the kernel-based effective memory $\Neff$ and summarise the implications of temporal averaging for collective order and responsiveness.

\section{Model and numerical protocol}
\label{sec:model}

\subsection{Finite buffer memory (FBM) model}

We consider $N$ self-propelled particles moving at constant speed $v_0$ in a two-dimensional periodic square of side $L$, with density $\rho=N/L^2$. Particle $i$ has position $\bm r_i$ and unit heading vector
\begin{equation}
\hat{\bm v}_i(t)=\left(\cos\theta_i(t),\sin\theta_i(t)\right).
\end{equation}
Particles within a metric interaction radius $R$ form the neighborhood $\mathcal N_i$, including particle $i$ itself. The instantaneous local alignment vector is
\begin{equation}
\bm V_i^{\mathrm{loc}}(t)=\sum_{j\in\mathcal N_i}\hat{\bm v}_j(t).
\end{equation}

Sensory uncertainty is represented by an angular error
\begin{equation}
\xi_i(t)\sim U(-\eta\pi,\eta\pi),
\end{equation}
where $0\leq\eta\leq1$ controls the noise amplitude. This is an \emph{angular} noise prescription: the direction of the local alignment vector is rotated, but its magnitude is not changed by the noise draw. It differs from adding an independent noise vector (often called vectorial or extrinsic noise); the two prescriptions need not produce identical ordering behavior~\cite{Gregoire2004}. The perceived local vector is
\begin{equation}
\widetilde{\bm V}_i^{\mathrm{loc}}(t)=
\mathcal R[\xi_i(t)]\bm V_i^{\mathrm{loc}}(t),
\label{eq:noisyvector}
\end{equation}
where $\mathcal R$ is the two-dimensional rotation matrix. Its angle,
$\widetilde\theta_i(t)=\arg(\widetilde{\bm V}_i^{\mathrm{loc}}(t))$, is inserted into a newest-in, oldest-out temporal memory buffer.

Unless otherwise stated, the noise-scan implementation stores the current observation and 20 previous observations, giving $M=21$ entries. In the separate fixed-noise deep-memory studies of Secs.~\ref{sec:high_noise_order} and~\ref{sec:relaxation_time}, $M$ is varied at $K=0$ rather than held at 21. The heading used in the subsequent step is the exponentially weighted circular mean
\begin{equation}
\theta_i(t+\Delta t)=
\arg\left[
\sum_{m=0}^{M-1}e^{-Km}e^{i\widetilde\theta_i(t-m\Delta t)}
\right].
\label{eq:buffer}
\end{equation}
The memory-decay parameter $K\geq0$ controls the temporal depth. At $K=0$, all stored observations have equal weight; as $K$ becomes large, the newest observation dominates, and the rule approaches a memoryless Vicsek-like update.

Positions are advanced according to
\begin{equation}
\bm r_i(t+\Delta t)=\bm r_i(t)+v_0\hat{\bm v}_i(t+\Delta t)\Delta t,
\end{equation}
with periodic boundary conditions. The displacement uses the updated heading $\hat{\bm v}_i(t+\Delta t)$, rather than the heading at time $t$.


\subsection{Observables}

The principal observable is the polar order parameter
\begin{equation}
\Psi(t)=\frac{1}{N}\left|\sum_{i=1}^{N}e^{i\theta_i(t)}\right|,
\label{eq:psi}
\end{equation}
where $\Psi\simeq1$ denotes the ordered state in which all agents share the same heading direction, and $\Psi\simeq0$ denotes a disordered state with approximately isotropic headings. Definitions and numerical analysis of additional fluctuation-based finite-size diagnostics are collected in Appendix~\ref{app:fluctuations}.

\subsection{Recursive learning model (RLM)}
\label{sec:ema}

To test whether the order enhancement is tied specifically to the explicit FBM architecture, we also consider a recursive learning model (RLM). Here, ``learning'' denotes the sequential incorporation of new noisy directional observations into a persistent internal state; there is no parameter training, objective-function minimization, or reinforcement-learning step. The noisy local vector is generated as in Eq.~(\ref{eq:noisyvector}), and the state $\bm H_i$ is updated as
\begin{equation}
\bm H_i(t+\Delta t)=(1-\alpha)\bm H_i(t)
+\alpha\widetilde{\bm V}_i^{\mathrm{loc}}(t),
\label{eq:ema}
\end{equation}
followed by $\theta_i(t+\Delta t)=\arg(\bm H_i(t+\Delta t))$. The persistence parameter $\alpha$ is restricted to the interval $0<\alpha\leq1$. A smaller $\alpha$ yields stronger persistence, while $\alpha=1$ corresponds to an instantaneous adjustment to current observations. The singular limit $\alpha=0$ freezes the initialized internal state and is not evaluated as a flocking controller. The recursion is an exponential moving average of noisy local alignment vectors. The direct algorithmic comparison is restricted to the common size $N=1600$ in Fig. \ref{fig:direct_comp}. The direct $\Psi(\eta)$ comparison uses $\alpha=0.1,0.2,0.4,0.6,0.8,$ and $1.0$, with noise sampled in increments of 0.05; the separate $\eta=0.8$ scans extend to smaller $\alpha$, including $1/\alpha$ up to $10^5$ in the insets of Figs.~\ref{fig:order_parameter} and~\ref{fig:relaxation_time}. 

\subsection{A common measure of memory depth}
\label{sec:neff}

For a normalized temporal kernel $w_m$, we define
\begin{equation}
\Neff\equiv\frac{1}{\sum_m w_m^2}.
\label{eq:neff_main}
\end{equation}
Under independent, sufficiently small angular observation errors, Eq.~(\ref{eq:neff_main}) represents the number of equally weighted observations that would produce the equivalent reduction in error variance. For the FBM model,
\begin{equation}
\Neff^{\rm buf}=
\frac{\left(\sum_{m=0}^{M-1}e^{-Km}\right)^2}
{\sum_{m=0}^{M-1}e^{-2Km}},
\label{eq:neffbuf_main}
\end{equation}
whereas the asymptotic RLM kernel gives
\begin{equation}
\Neff^{\rm RLM}=\frac{2-\alpha}{\alpha}.
\label{eq:neffema_main}
\end{equation}
This expression characterizes the weights of the temporal filter. Its interpretation as an independent-sample count assumes weak and uncorrelated angular errors. Appendix~\ref{app:memory} gives both the small-error derivation and an exact circular-noise calculation for a fixed underlying direction. For $M=21$, setting $K=0.05$ yields $\Neff^{\rm buf}\simeq19.27$, closely matching $\alpha=0.1$, which yields $\Neff^{\rm RLM}=19$.

\subsection{Simulation parameters} 
\label{sec:simulation_parameters} 
Unless stated otherwise, all simulations are performed at particle speed $v_0=0.5$, interaction radius $R=1$, time step $\Delta t=0.1$ and number density $\rho=4$. System sizes range from $N=400$ to $N=6400$. For each parameter set, observables are averaged over $50$ independent realisations. Each realisation is evolved for $T=10\,000$ time units, with the first $5000$ time units discarded as transient. The reported observables are obtained by first computing the ensemble average over independent realisations and subsequently averaging over the post-transient time window. For the principal noise scans of the FBM, we use system sizes $N=1600$, $3600$, and $6400$. The finite-size polarisation analysis in Fig.~\ref{fig:fss} additionally includes smaller systems and displays results for $N=400$, $900$, $1600$, and $6400$. The sampled decay parameters are \[ K\in\{0,\,0.005,\,0.01,\,0.05,\,0.1,\,0.2,\,0.5,\,2,\,5\}. \] For the fixed buffer length $M=21$, the oldest and most recent observations are separated by \[ (M-1)\Delta t=2.0. \] Over this interval, a particle moving at $v_0=0.5$ travels a distance \[ v_0(M-1)\Delta t=1.0, \] which is equal to the interaction radius $R=1$. This comparison does not provide a direct measure of neighbour-turnover times, but it indicates that consecutive observations stored in the buffer need not be statistically independent. For the RLM, the learning parameter is varied over the range $0\leq\alpha\leq0.1$. The remaining numerical parameters and averaging procedure are unchanged. Simulations with $\alpha<0.01$ are instead run for $T=50\,000$ time units with first $25000$ time units discarded as transient to accommodate the longer transient associated with increasingly persistent memory. For all simulations, initially, particles were distributed uniformly at random within a 2D box of length $L$, and standard periodic boundary conditions were applied. The entries of each FBM memory buffer were initially set to zero angle, and each RLM internal vector was initialized as $H_i(0)=0$. Initial particle headings were independently taken randomly from uniform distribution $[-\pi,\pi)$.

\section{Results}
\label{sec:results}

\subsection{Memory shifts the loss of collective order to higher noise}

\begin{figure}[tb]
\centering
\includegraphics[width=\columnwidth]{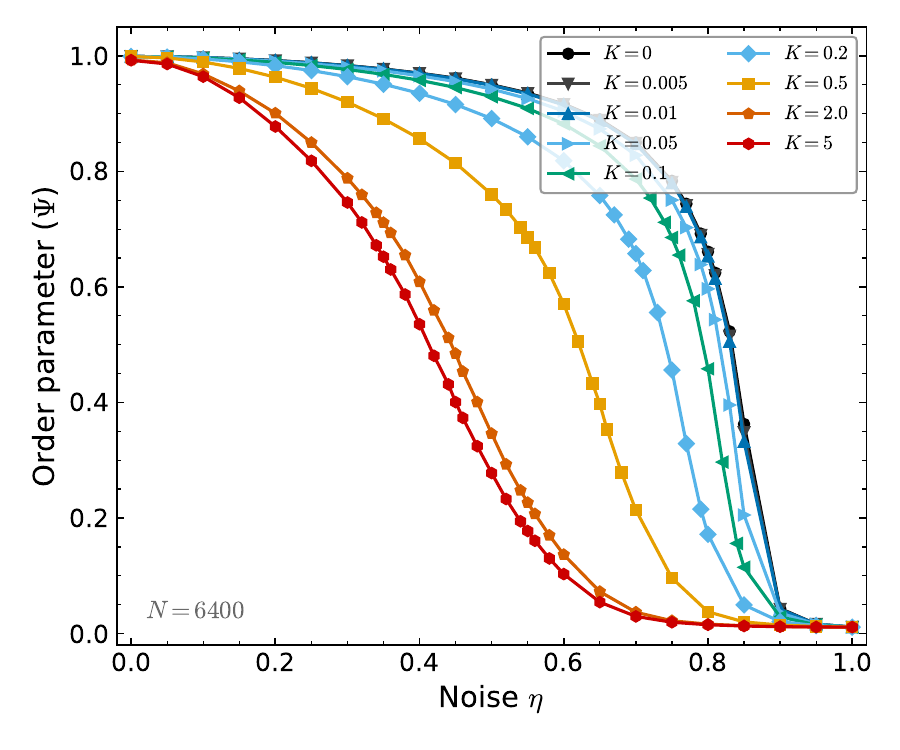}
\caption{Steady-state polarization $\Psi$ versus angular-noise amplitude $\eta$ at $N=6400$ for the nine FBM decay values $K=0$, $0.005$, $0.01$, $0.05$, $0.1$, $0.2$, $0.5$, $2$, and $5$. For $M=21$, the four smallest decay rates correspond to a narrow range $\Neff\simeq19.27$--$21$, which accounts for their closely spaced curves. Lower $K$ weights older observations more strongly and moves the loss-of-order region toward larger $\eta$; $K=5$ is close to the short-memory limit for the fixed $M=21$ buffer.}
\label{fig:op}
\end{figure}

Figure~\ref{fig:op} illustrates the principal result. At a fixed noise amplitude within the loss-of-order region, decreasing $K$ consistently produces a larger steady-state polarization. Using the finite-system midpoint $\eta_{50}$ defined by $\Psi(\eta_{50})=0.5$ and determined by linear interpolation on the sampled grid [Fig.~\ref{fig:collapse}(a)], the short-memory limit $K=5$ has $\eta_{50}\simeq0.45$, compared with $\eta_{50}\simeq0.837$ for $K=0$. The midpoint thus increases by approximately $86\%$ (a factor of $1.86$) over this scan. The descent occurs in broad ranges around these midpoints rather than at sharply defined critical noise values. In particular, a curve with $\Psi\approx0.6$ at $\eta=0.8$ should not be described as nearly perfectly ordered.

This trend is continuous across the nine sampled values of $K$: weighting earlier observations more strongly shifts the polarization decrease toward higher noise. At $\eta=0.7$, for example, the long-memory curves retain considerably more order than the nearly memoryless case. Throughout this work, $\eta_{50}$ is a descriptive midpoint for the measured finite systems, not a thermodynamic critical parameter.

\begin{figure}[tb]
\centering
\includegraphics[width=\columnwidth]{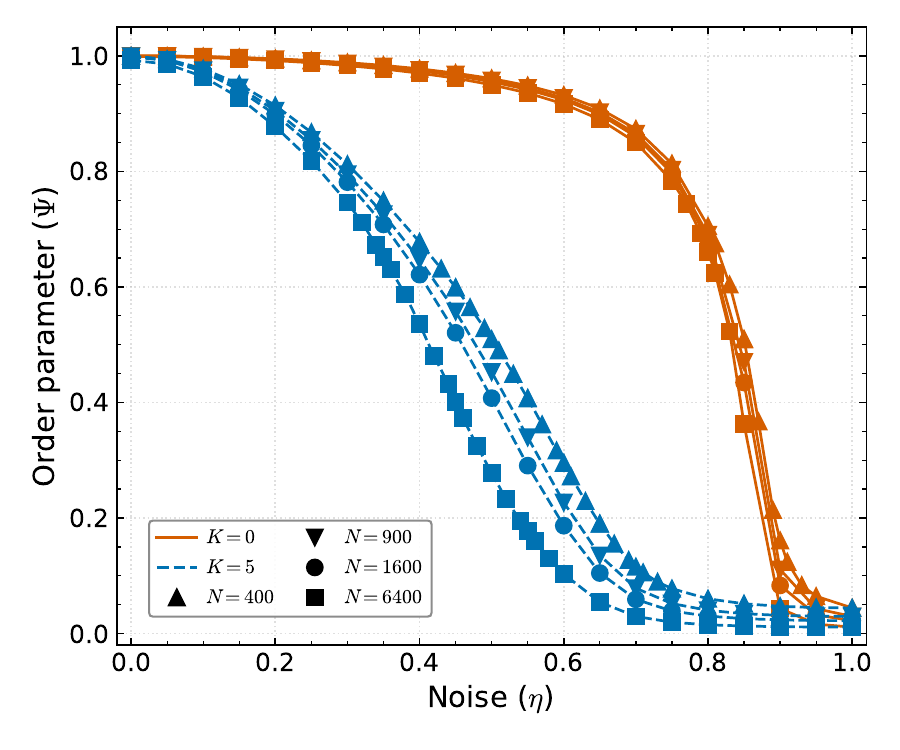}
\caption{Finite-size polarization $\Psi(\eta)$ in the FBM for $K=0$ (solid orange) and $K=5$ (dashed blue), at the four sizes actually shown: $N=400$, $900$, $1600$, and $6400$ (different symbols). The two memory choices have well-separated loss-of-order regions throughout this sampled size range; the size dependence is not used to extract an asymptotic critical point.}
\label{fig:fss}
\end{figure}

Figure~\ref{fig:fss} provides a direct finite-size comparison for two extreme $K$ values evaluated in this study. Across the four sizes displayed ($N=400,900,1600,6400$), increasing $N$ generally sharpens the decline of $\Psi$ and modestly shifts the finite-size loss-of-order region; however, it does not eliminate the wide separation between the $K=0.0$ and $K=5$ noise scales. The memory-related separation is therefore visible at every plotted size. The figure does not report $\eta_{50}(N)$ values with uncertainties or resolve a common Binder crossing, so its size dependence should not be interpreted as a quantitative finite-size scaling law. A substantially larger and more finely resolved finite-size scaling analysis would be required to robustly extrapolate the asymptotic transition location or define its universal order. Supporting finite-size fluctuation diagnostics are presented separately in Appendix~\ref{app:fluctuations} (Figs.~\ref{fig:binder} and~\ref{fig:chi}).

\subsection{Direct comparison between FBM and RLM}

\begin{figure}[tb]
\centering
\includegraphics[width=\columnwidth]{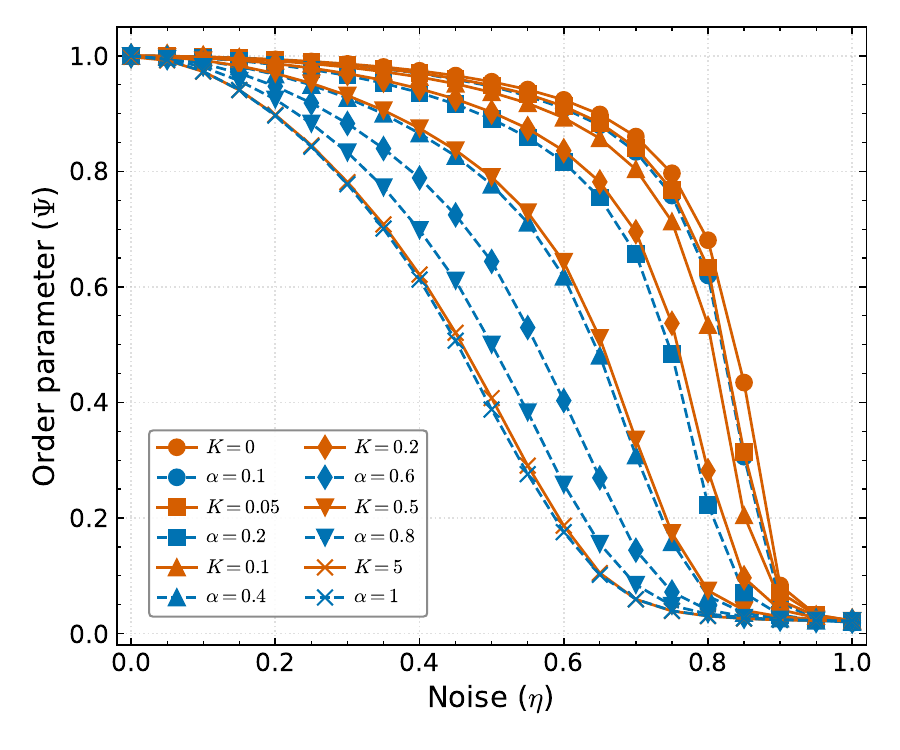}
\caption{Direct comparison of steady-state $\Psi(\eta)$ for the FBM (solid orange; varying $K$) and RLM (dashed blue; varying $\alpha$) at $N=1600$. Lowering $K$ or $\alpha$ increases temporal persistence and shifts the fall of polarization toward higher noise. The RLM set does not include the singular frozen-state limit $\alpha=0$.}
\label{fig:direct_comp}
\end{figure}

We next compare the FBM with the recursive update rule to examine whether the stationary trend depends on storing an explicit history. Figure~\ref{fig:direct_comp} presents a direct comparison of the steady-state polar order parameter $\Psi$ as a function of the noise amplitude $\eta$ for both memory implementations. The solid orange lines plot the FBM model across various decay parameters $K$, while the dashed blue lines represent the RLM across different persistence parameters $\alpha$. The plotted parameter sets reflect separate scans and are not identical. In both models, decreasing $K$ or $\alpha$ moves the drop in polarization to higher noise. The FBM averages angles stored in a finite buffer, whereas the RLM updates a persistent vector. Similar steady-state behavior is therefore obtained from two different memory rules.

The curves also suggest that $\Neff$ is a useful scale for comparing the two filters. In particular, the RLM curve for $\alpha=0.1$ ($\Neff=19$) closely follows the FBM curve for $K=0.05$ ($\Neff\simeq19.27$) across the sampled noise range. At $\alpha=1$ the RLM has no recursive persistence, while the $K=5$, $M=21$ FBM kernel has $\Neff\simeq1.014$ and is nearly instantaneous. Their plotted profiles provide a useful qualitative short-memory comparison, but were not evaluated under fully matched sampling protocols. This comparison concerns the stationary polarization, not the full dynamics.

\subsection{High-noise order enhancement via deep memory}
\label{sec:high_noise_order}

\begin{figure}[tb]
\centering
\includegraphics[width=\columnwidth]{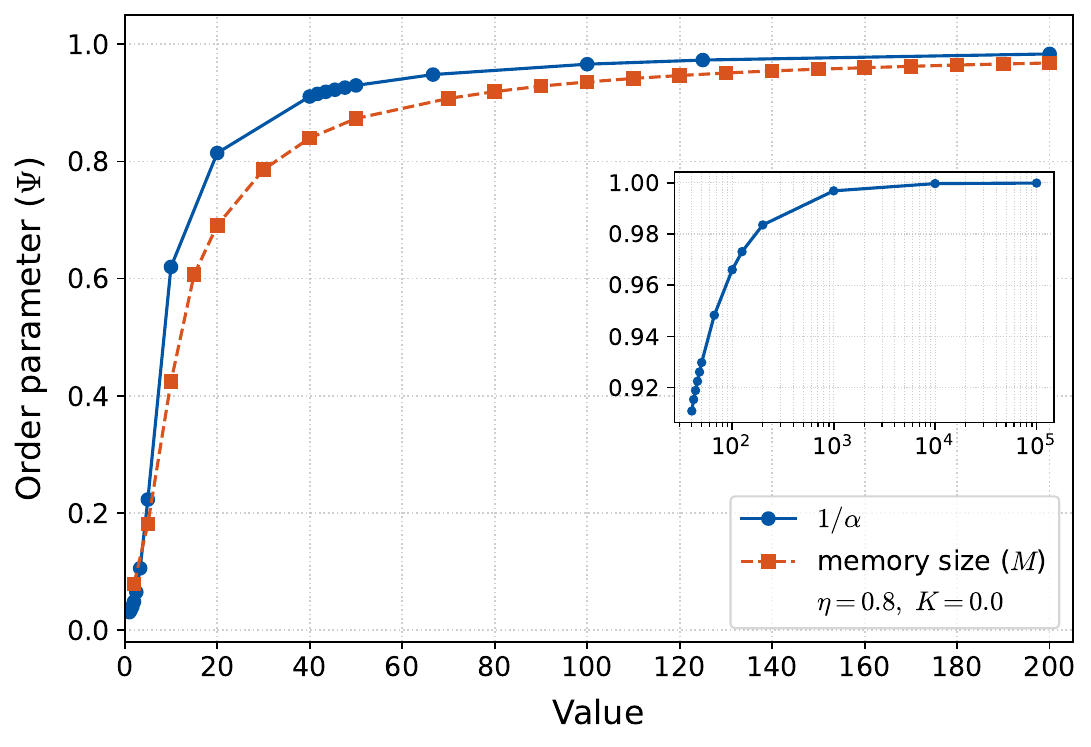}
\caption{Steady-state polarization $\Psi$ at fixed $\eta=0.8$ as the RLM inverse update fraction $1/\alpha$ increases (blue circles, solid line) or the uniformly weighted FBM buffer length $M$ increases at $K=0$ (orange squares, dashed line). Both curves rise and exceed $\Psi\simeq0.9$ at sufficiently deep memory; the RLM reaches $\Psi\simeq0.98$ by $1/\alpha\simeq200$ and approaches unity over the wider range shown in the logarithmic-$1/\alpha$ inset (up to $10^5$). The shared horizontal axis denotes two different parameters, not equal effective-memory depths: at $K=0$, $\Neff^{\rm buf}=M$ while $\Neff^{\rm RLM}=2/\alpha-1$.}
\label{fig:order_parameter}
\end{figure}

The fixed-buffer noise scans above vary the decay rate $K$ at $M=21$. To examine substantially longer temporal histories, we now fix the strong angular-noise amplitude at $\eta=0.8$ and vary the persistence of each implementation independently. For the FBM, we increase $M$ with uniform weights ($K=0$); for the RLM, we increase the inverse update fraction $1/\alpha$. Figure~\ref{fig:order_parameter} shows that both changes increase the stationary polarization over the sampled parameter ranges. Near the short-memory end of the sweep, the order parameter is small. It rises to about $0.9$ at $1/\alpha\sim40$ for the RLM and at a buffer length of order $M\sim70$--$100$ for the FBM. At $1/\alpha\simeq200$ and $M=200$, the respective values are approximately $0.98$ and $0.97$, as read from the plot. In the RLM inset, $\Psi$ approaches unity as $1/\alpha$ is increased to $10^4$--$10^5$.

The shared horizontal axis in Fig.~\ref{fig:order_parameter} represents different parameters in the two models. At $K=0$, the normalized FBM kernel gives $\Neff^{\rm buf}=M$; for the RLM, Eq.~(\ref{eq:neffema_main}) gives $\Neff^{\rm RLM}=2/\alpha-1=2(1/\alpha)-1$. Thus the RLM has roughly twice the kernel-based effective depth at equal, sufficiently large horizontal-axis values. This difference accounts for part of the horizontal offset. The filter weights alone do not predict polarization in the interacting flock.

\subsection{Relaxation dynamics and optimal memory for order recovery}
\label{sec:relaxation_time}

The steady-state trend does not determine how quickly a flock can regain collective order after a disturbance. We therefore define a recovery or first-passage time $\tau$ as the number of simulation time needed to return to $\Psi>0.9$ after the polarization has been reduced to approximately $\Psi=0.5$ at $\eta=0.8$. The measured $\tau$ is a threshold-crossing time, not a relaxation constant obtained from an exponential fit.

To measure recovery from a perturbation of the internal memory state, each realisation is first evolved until $\Psi>0.9$. Particle positions are then held fixed while independent angular perturbations of amplitude $\eta=0.1$ are applied repeatedly to the persistent vectors $\bm{H}_i$ in the RLM or to the stored angular entries in the FBM. The perturbation of a generic vector $\bm{v}$ is defined by \[ \bm{v}\longrightarrow\mathcal{R}[\xi_i]\bm{v}, \qquad \xi_i\sim\mathcal{U}(-\eta\pi,\eta\pi), \] where $\mathcal{R}[\xi_i]$ is the rotation operator through the angle $\xi_i$. For the FBM, the operator is applied to the unit vector associated with each stored angle, and the resulting direction is written back to the corresponding buffer entry. Headings and polarisation are recomputed after each perturbation step, without updating the particle positions, until $0.48<\Psi<0.52$. The full coupled dynamics is then restored, and the relaxation time $\tau$ is defined by the first subsequent crossing of $\Psi>0.9$. Means and standard deviations are computed from $50$ independent realisations for each parameter set.

\begin{figure}[tb]
\centering
\includegraphics[width=\columnwidth]{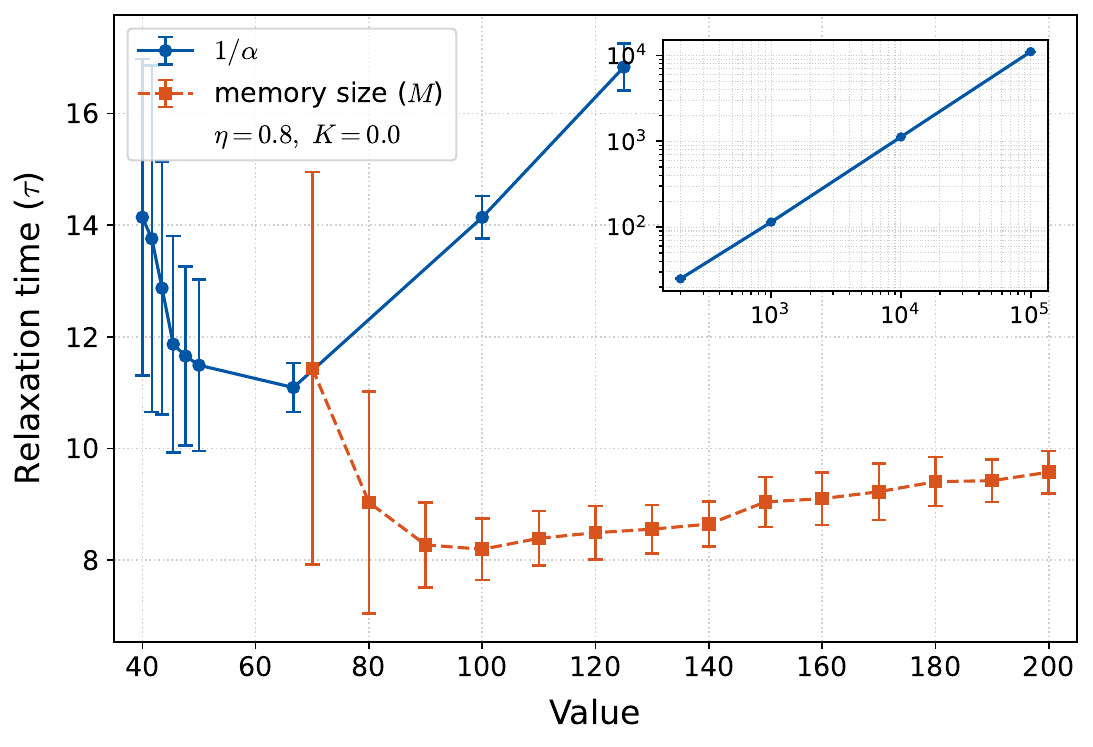}
\caption{Mean first-passage recovery time $\tau$ to reach $\Psi>0.9$ after perturbation to $\Psi\simeq0.5$, at $\eta=0.8$. Blue circles (solid line) show the RLM versus $1/\alpha$; orange squares (dashed line) show the FBM versus $M$, with $K=0$ applying only to the FBM. Error bars are reported as standard deviations across 50 realizations. The RLM has a sampled minimum around $1/\alpha\simeq67$ ($\tau\simeq11.1$); the FBM reaches a shallow minimum at $M=100$ ($\tau\simeq8.2$), within a broad region of overlapping error bars. The RLM inset uses logarithmic axes and displays a roughly proportional rise of $\tau$ with $1/\alpha$ over its largest sampled values, without a fitted exponent. Equal horizontal-axis values do not correspond to equal $\Neff$.}
\label{fig:relaxation_time}
\end{figure}

Figure~\ref{fig:relaxation_time} shows a nonmonotonic recovery-time trend over the sampled ranges, in contrast to the rising steady-state polarization of Fig.~\ref{fig:order_parameter}. In the RLM, the measured $\tau$ decreases from approximately $14$ steps near $1/\alpha=40$ to about $11.1$ steps at $1/\alpha\simeq67$, before rising to roughly $17$ steps at $1/\alpha\simeq125$. In the FBM, the initial point at $M=70$ gives $\tau\simeq11.4$ steps with a large dispersion. Recovery becomes faster with increasing $M$ and reaches a minimum at $M=100$, where $\tau\simeq8.2$ steps, within a broad low-recovery-time region; by $M=200$, it rises gradually to approximately $9.6$ steps. Because the FBM error bars overlap over a range of $M$, the data support a favorable intermediate interval centered around $M=100$ rather than a statistically established unique optimum.

The RLM inset extends the persistence sweep beyond the main panel. On its logarithmic axes, the points at $1/\alpha\sim10^3$, $10^4$, and $10^5$ lie close to a slope-one trend, with $\tau$ increasing from order $10^2$ to order $10^4$ steps. No fitted exponent or uncertainty estimate is supplied for this apparent scaling. A plausible mechanism for the late-time slowdown is the greater persistence of perturbed headings in the memory state: filtering reduces noisy fluctuations but also limits the rate of readjustment when the desired collective direction changes. This interpretation is consistent with the plotted results, although the relative roles of internal-state memory, collective feedback, and the chosen perturbation protocol remain to be tested directly. As for the stationary plot, equal values of $1/\alpha$ and $M$ do not imply equal $\Neff$ or an equal dynamical response.

\subsection{Supporting robustness test with a second memory rule}
\label{sec:comparison}

\begin{figure*}[tb]
\centering
\includegraphics[width=0.94\textwidth]{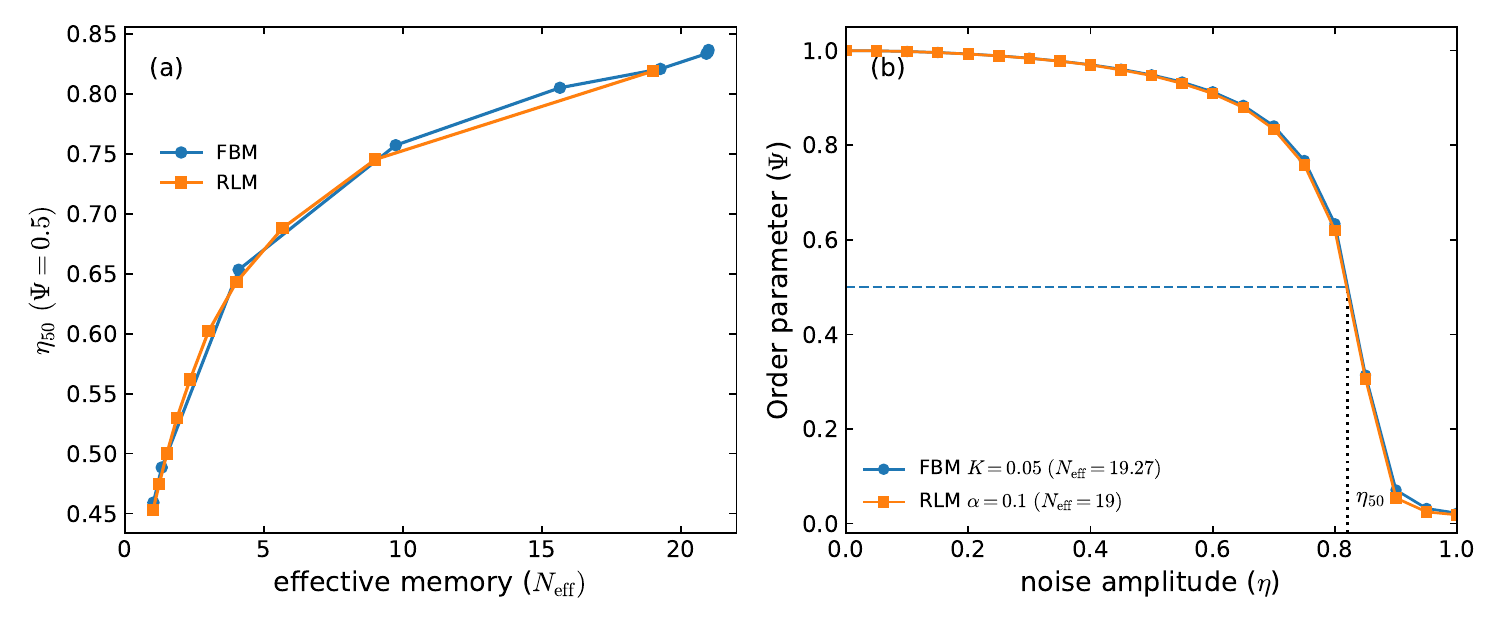}
\caption{Comparison of the two memory schemes in terms of kernel-based effective memory $\Neff$. (a) The finite-size operational noise midpoint $\eta_{50}$, obtained by interpolating $\Psi(\eta)=0.5$, versus $\Neff$ for FBM and RLM. (b) Steady-state polarization versus $\eta$ for FBM at $K=0.05$ ($\Neff\simeq19.27$) and RLM at $\alpha=0.1$ ($\Neff=19$). The horizontal dashed line marks $\Psi=0.5$ and the vertical dotted line indicates the corresponding approximate $\eta_{50}$. This is a comparison of steady-state behavior, not a demonstration of dynamical equivalence.}
\label{fig:collapse}
\end{figure*}

Figure~\ref{fig:collapse}(a) summarizes the cross-implementation comparison in terms of effective memory depth. We use the operational noise midpoint $\eta_{50}$, defined in Sec.~\ref{sec:results} by linearly interpolating to $\Psi=0.5$ on each plotted $\Psi(\eta)$ curve [Fig.~\ref{fig:collapse}(b)]. Over the sampled values of $\Neff$, the FBM and RLM thresholds follow similar monotonic trends. In the short-memory regime, $\eta_{50}$ rises from approximately $0.45$ near $\Neff\simeq1$ to approximately $0.65$ around $\Neff\simeq4$ (for example, $K=0.5$ gives $\Neff\simeq4.08$ and $\eta_{50}\simeq0.653$, whereas $\alpha=0.4$ gives $\Neff=4$ and $\eta_{50}\simeq0.643$). The increase becomes more gradual at larger $\Neff$: extending the effective depth from roughly $4$ to $20$ produces a smaller additional displacement than the initial increase from $1$ to $4$. This trend motivates the circular-noise estimate in Appendix~\ref{app:memory}, but neither $\eta_{50}\propto\sqrt{\Neff}$ nor a universal threshold collapse follows from the present data. At the largest depths sampled, $\eta_{50}$ reaches approximately $0.82$--$0.84$ ($K=0$: $\eta_{50}\simeq0.837$; $K=0.05$: $\eta_{50}\simeq0.821$; $\alpha=0.1$: $\eta_{50}\simeq0.819$).

A direct comparison at nearly matched memory depth is shown in Fig.~\ref{fig:collapse}(b) for the FBM with $K=0.05$ ($\Neff\simeq19.27$) and the RLM with $\alpha=0.1$ ($\Neff=19$). Their steady-state polarization profiles are very close on the sampled noise grid through the loss-of-order region, not only at the operational midpoint $\eta_{50}$. This agreement is nontrivial because the FBM averages stored unit phasors, whereas the RLM recursively updates a state vector whose magnitude depends on local neighbor number and coherence. The close stationary curves are consistent with noise filtering by both memory rules, despite their distinct microscopic updates.

\section{Discussion}
\label{sec:discussion}

The simulations show that the temporal integration stabilises collective motion by averaging over fluctuations in the perceived local direction. When the collective state evolves slowly compared with the memory timescale, this averaging produces a more reliable alignment signal and allows polar order to persist under stronger angular noise. The calculation in Appendix~\ref{app:memory} supports this filtering picture, although it does not constitute a quantitative theory of the interacting system: successive observations are correlated, and the local reference direction evolves together with the flock.

A central result is that the stationary responses of the FBM and RLM are organised reasonably well by the kernel-based effective memory depth $\Neff$ (Fig.~\ref{fig:collapse}). This correspondence is notable because the two models store and update directional information differently. It suggests that stationary ordering is governed primarily by the temporal weighting of past observations, while the details of the memory implementation are less important. The match should nevertheless be regarded as an empirical correspondence rather than evidence that the two microscopic dynamics are equivalent.

The recovery dynamics presents a different picture (Fig.~\ref{fig:relaxation_time}). In both models, the relaxation time is non-monotonic in memory depth. Short memories provide insufficient protection against angular noise, whereas very persistent memories retain perturbed directional information and delay collective realignment. Recovery is therefore fastest at an intermediate memory depth, where noise suppression is balanced against the replacement of outdated information. In the RLM, recovery becomes progressively slower as $\alpha$ decreases. The data at large $1/\alpha$ are approximately consistent with a slope-one trend on logarithmic axes, but they are not sufficient to establish an asymptotic scaling law.

The different recovery curves of the FBM and RLM can be traced to the support of their memory kernels. The FBM removes an observation after a finite number of updates, whereas the RLM retains an exponentially decaying contribution from earlier states. This distinction has little effect on stationary ordering but becomes important when the internal memory is itself perturbed. A finite buffer clears the corrupted history within a bounded time, while a recursive memory retains a progressively weaker remnant of it.

This interpretation is specific to the perturbation considered here. Because both the instantaneous headings and the internal memory variables are perturbed, the measured relaxation describes recovery from a corrupted directional history. A perturbation applied only to the current headings, while leaving the memory intact, produces recovery in a single time step as the particles average the whole memory and align with the resulting direction regardless of the current heading. This picture will differ for shallow memory, where the current observation dominates past observations.

The effective depth $\Neff$ should not be interpreted as an exact number of independent observations. Consecutive samples are correlated through overlapping interaction neighbourhoods, and the underlying collective direction is not fixed. These effects, particularly at high noise, lie outside the independent-sample argument used in Appendix~\ref{app:memory}. A quantitative prediction of the ordering crossover would require a kinetic description coupling memory to density, local polarisation, and the evolving interaction network.

The present mechanism also differs from inertial and delayed-response models of collective motion~\cite{Ballerini2008,Cavagna2010,Bialek2012,HerbertRead2011,Katz2011,Attanasi2014,Cavagna2015,Nagai2015,Forgoston2008,Durve2018}. Here, memory acts primarily as a low-pass filter: it suppresses rapidly varying angular errors but also reduces sensitivity to genuine changes in the collective direction. For decentralised swarms and multi-agent systems, the relevant design objective is therefore not maximal memory, but a memory timescale that balances noise resilience against responsiveness. Finite-horizon filters may be advantageous when obsolete information must be removed within a bounded time.

\section{Conclusion}
\label{sec:conclusion}

We have studied how temporal integration of noisy local information affects both the stationary ordering and the transient response of self-propelled particles. To separate robust features from those tied to a particular implementation, we considered two forms of memory: a finite buffer of previous observations (FBM) and a recursively learned state (RLM). In both cases, memory extends polar order into a regime in which the corresponding memoryless Vicsek system is disordered. Despite their different microscopic definitions, the stationary responses of the two models can be compared through the kernel-based effective memory depth, $\Neff$. This matching suggests that the improvement in stationary order is controlled primarily by the temporal averaging of angular fluctuations, rather than by the particular mechanism used to retain past information.

The response to a collective perturbation is less straightforward. Increasing the memory depth improves stationary polarisation at high noise, but it does not lead to progressively faster recovery. Instead, the relaxation time $\tau$ is non-monotonic, with the most rapid recovery occurring at an intermediate memory depth. Beyond this point, the information retained from earlier states becomes increasingly difficult to overwrite, and collective realignment slows down. In the persistent-memory regime of the RLM, this slowdown is approximately linear in the persistence depth, with $\tau \propto 1/\alpha$. Memory therefore plays two competing roles. It filters short-lived angular fluctuations, thereby stabilising the ordered state, but it also retains directional information that becomes obsolete after a sufficiently large perturbation. The memory depth that gives the strongest stationary order is consequently not, in general, the one that gives the fastest dynamical response.

This distinction is relevant to decentralised swarms and multi-agent systems operating with noisy local measurements. Temporal filtering can improve coherence without requiring global information, but an overly persistent filter can compromise manoeuvrability when the collective direction must change. The appropriate memory depth should therefore be chosen with the expected operating conditions in mind: systems required to maintain a fixed course can benefit from stronger temporal averaging, whereas systems exposed to abrupt changes or external commands require a shorter effective memory.

Several questions remain open. The simulations reported here characterise the ordering and recovery of finite systems, but they do not establish how temporal memory affects the asymptotic flocking transition. Addressing this issue will require finite-size scaling together with a systematic analysis of density inhomogeneities and travelling bands. A kinetic description would also be useful for determining how the memory kernel enters the coarse-grained dynamics and how it couples to the evolving interaction network. In particular, such a theory would need to account for correlations between internal memory states, local density, and the changing neighbourhoods sampled by moving particles. Extending the present models to time-dependent external fields, spatially heterogeneous noise, and non-stationary environments should clarify whether the same compromise between robustness and responsiveness persists under more general conditions.

\begin{acknowledgments}
The authors acknowledge IIT's HPC Franklin for carrying out the simulations. The authors acknowledge the use of AI to assist with drafting manuscript text and developing research code. 
\end{acknowledgments}

\appendix
\setcounter{figure}{0}
\renewcommand{\thefigure}{A\arabic{figure}}
\providecommand{\theHfigure}{\thefigure}
\renewcommand{\theHfigure}{appendix.A.\arabic{figure}}
\section{Finite-size fluctuation diagnostics}
\label{app:fluctuations}

The main text uses the polarization $\Psi$ and its particle-number dependence to characterize the observed loss of collective order. Here we report two complementary finite-size observables and the accompanying plots. These diagnostics characterize the sampled finite systems; they are not used to determine a thermodynamic transition point or universality class.

\subsection{Definitions}

We evaluate the Binder cumulant~\cite{Binder1981}, a common finite-size diagnostic~\cite{Binder1997},
\begin{equation}
G=1-\frac{\avg{\Psi^4}}{3\avg{\Psi^2}^2},
\label{eq:binder}
\end{equation}
and the susceptibility-like measure of polarization fluctuations,
\begin{equation}
\chi=N\left(\avg{\Psi^2}-\avg{\Psi}^2\right).
\label{eq:chi}
\end{equation}
The brackets denote empirical moments of the sampled polarization, rather than fluctuations of individual particle headings. Precisely how the time and realization averages were pooled is an implementation detail that should be reported from the numerical analysis, not inferred from the plots. For independent isotropic headings, the large-$N$ disordered reference is $G\to1/3$ and, with the present normalization, $\chi\to1-\pi/4\simeq0.215$; finite sampling and temporal correlations can modify the observed values. The structures in $G$ and $\chi$ are used as finite-size diagnostics, not as exact critical coordinates.

\subsection{Binder cumulant}

\begin{figure}
\centering
\includegraphics[width=\columnwidth]{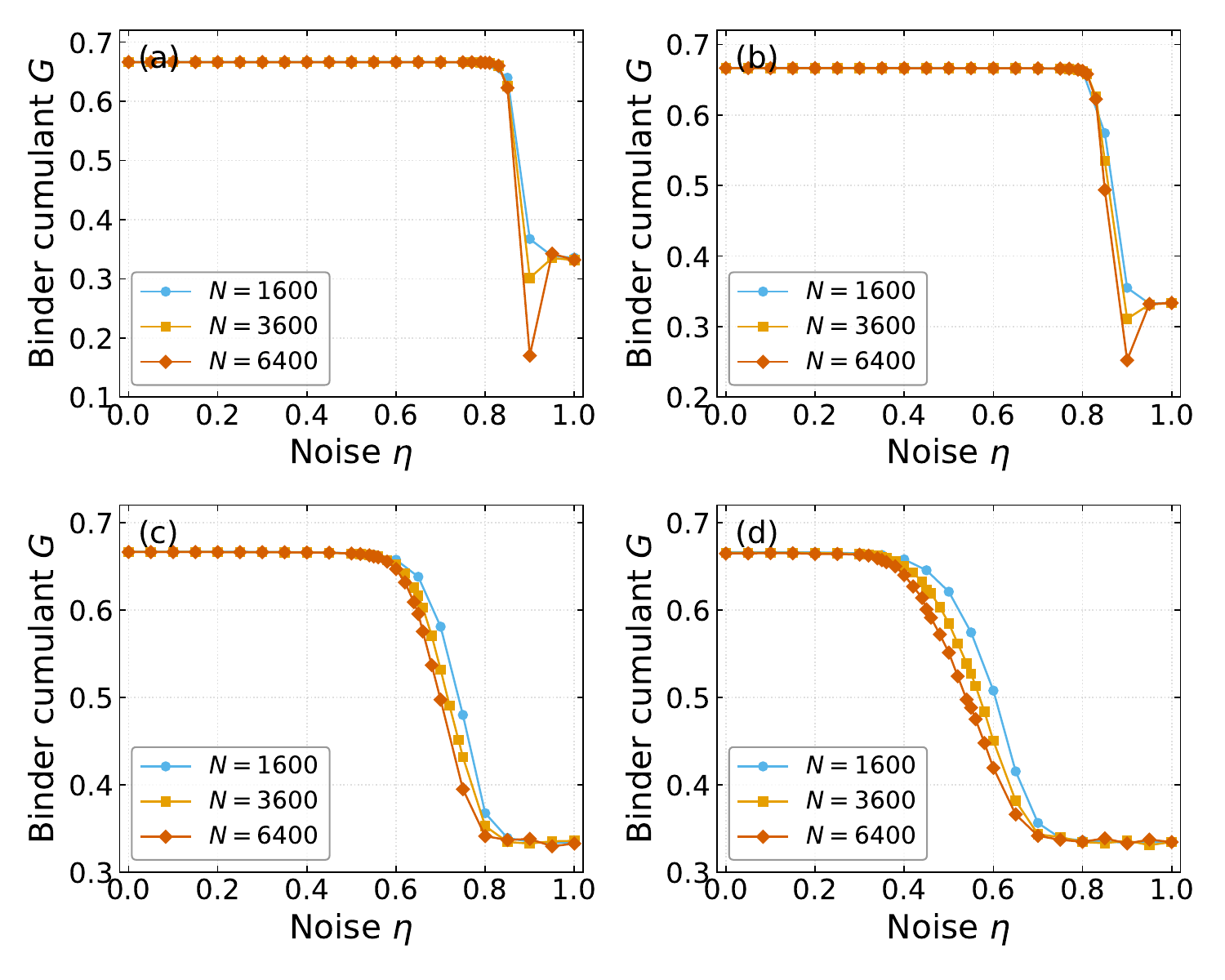}
\caption{Binder cumulant $G$ against noise $\eta$ for $N=1600$, $3600$, and $6400$ (symbols and colors as in the legends). Panels (a)--(d) display successively left-shifted loss-of-order regions. From panels (a) to (d), the changes in $G$ occur at progressively lower noise, with pronounced nonmonotonic dips in some of the high-noise panels. Minima near the loss-of-order region are finite-size signatures, not by themselves a classification of transition order.}
\label{fig:binder}
\end{figure}

The Binder cumulant presented in Fig.~\ref{fig:binder} remains near its ordered-state plateau throughout the low-noise regime, dropping sharply across the same qualitative noise range as the macroscopic polarization. In panels (a) and (b), whose rapid declines occur at relatively high noise, the larger system sizes exhibit a nonmonotonic minimum near the drop in $\Psi$. Such structural features are consistent with strong finite-size fluctuations and are frequently associated with first-order-like coexistence behaviors in Vicsek-type active matter models. Because these minima are represented by a sparse noise grid for three sizes, a common size-independent Binder crossing cannot be reliably identified from the plotted points. These curves therefore do not determine the asymptotic transition order.

\subsection{Susceptibility-like fluctuations}

\begin{figure}
\centering
\includegraphics[width=\columnwidth]{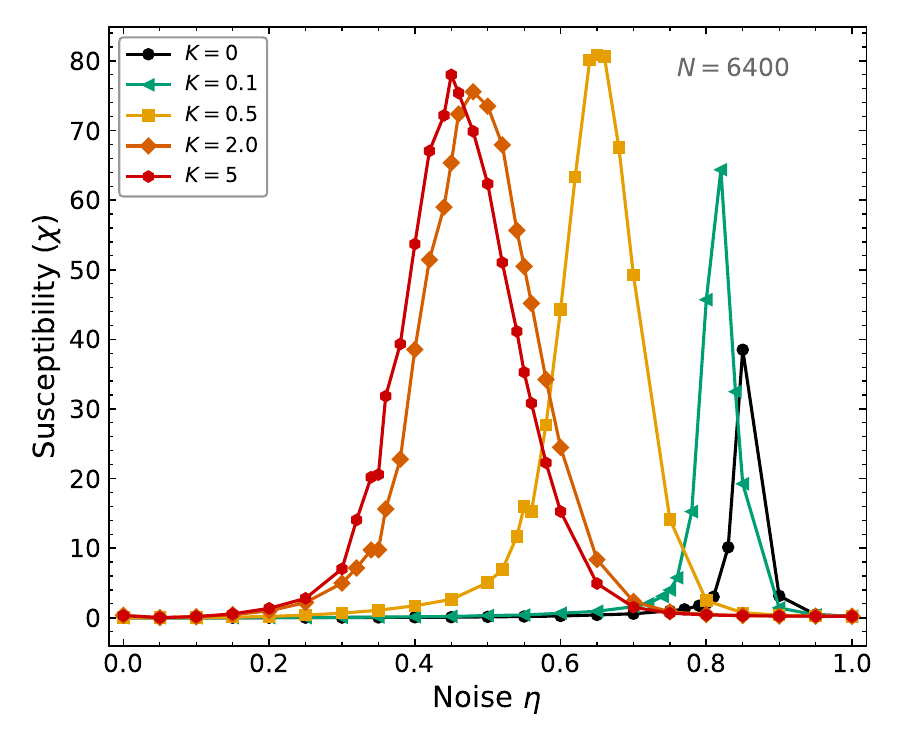}
\caption{Susceptibility-like fluctuations $\chi=N(\langle\Psi^2\rangle-\langle\Psi\rangle^2)$ versus $\eta$ at $N=6400$ for $K=0$, $0.1$, $0.5$, $2$, and $5$. The curves have comparable low baselines and distinct maxima: approximately $\eta\simeq0.45$--$0.5$ for $K=2,5$, $\eta\simeq0.65$ for $K=0.5$, and $\eta\simeq0.8$--$0.85$ for $K=0,0.1$. Peak positions are approximate because of the finite noise grid; their movement with $K$ is the main observation.}
\label{fig:chi}
\end{figure}

The fluctuation measure $\chi$ offers a complementary diagnostic perspective. Figure~\ref{fig:chi} demonstrates that the regime of maximal fluctuation shifts systematically toward larger $\eta$ as $K$ decreases. All five curves have comparable low-noise baselines, and their principal peaks shift in sequence from $K=5$ and $2$ near $\eta\simeq0.45$--$0.5$, through $K=0.5$ near $0.65$, to $K=0.1$ and $0$ near $0.8$--$0.85$. The shorter-memory curves retain broader shoulders around their maxima. Peak heights and widths also depend on numerical sampling and the fluctuation estimator.

\subsection{Scope of the finite-size interpretation}

The Binder cumulant and susceptibility-like fluctuations provide complementary finite-size diagnostics of the displaced order-loss region (Figs.~\ref{fig:binder} and~\ref{fig:chi}). In the corrected susceptibility curves, the maxima move systematically toward higher noise as memory is deepened; the low-noise baselines are comparatively similar across $K$. The nonmonotonic Binder minima may be compatible with strong finite-size or coexistence-like fluctuations, but interpreting these structures as evidence of the asymptotic transition order would exceed the available resolution. A thermodynamic assessment would require larger systems, more densely sampled noise values, order-parameter distributions, hysteresis tests, and spatial density-band diagnostics~\cite{Chate2008,Baglietto2009}. These observables should be interpreted together with the polarization curves rather than in isolation.

\section{Effective-memory calculation and limitations of the comparison}
\label{app:memory}

The purpose of this appendix is to map the FBM of Eq.~(\ref{eq:buffer}) and the recursive RLM of Eq.~(\ref{eq:ema}) onto a shared metric of temporal averaging capacity. This mapping does not presume the two algorithms are dynamically identical.

\subsection{General weighted-noise calculation}

Assuming that the underlying local direction remains approximately constant at $\theta_0$ across the relevant memory window, the observed noisy angles are given by
\begin{equation}
\widetilde\theta(t-m\Delta t)=\theta_0+\xi(t-m\Delta t),
\end{equation}
with $\avg{\xi_t}=0$ and $\mathrm{Var}(\xi_t)=\sigma^2$. For the uniform angular noise distribution utilized here, the variance is
\begin{equation}
\sigma^2=\frac{(\eta\pi)^2}{3}.
\end{equation}
For a normalized temporal kernel $w_m$ where $\sum_mw_m=1$, a small-error expansion yields
\begin{equation}
\sum_mw_me^{i\widetilde\theta(t-m\Delta t)}
\simeq e^{i\theta_0}\left(1+i\sum_mw_m\xi(t-m\Delta t)\right).
\end{equation}
The leading-order angular error in the filtered heading is therefore
\begin{equation}
\xi_{\mathrm{eff}}\simeq\sum_mw_m\xi(t-m\Delta t).
\end{equation}
If the temporal observation errors are strictly independent, the effective variance becomes
\begin{equation}
\mathrm{Var}(\xi_{\mathrm{eff}})=\sigma^2\sum_mw_m^2.
\label{eq:vareff}
\end{equation}
Because an equal-weight average across $n$ independent observations reduces the variance to $\sigma^2/n$, we define
\begin{equation}
\boxed{\Neff=\frac{1}{\sum_mw_m^2}}
\end{equation}
as the equivalent number of equally weighted, independent observations necessary to achieve the identical linearized variance reduction. For positive, unnormalized weights $a_m$, this generalizes to
\begin{equation}
\Neff=\frac{\left(\sum_ma_m\right)^2}{\sum_ma_m^2}.
\label{eq:neff_unnorm}
\end{equation}

If the observation errors possess temporal covariance, Eq.~(\ref{eq:vareff}) must incorporate off-diagonal terms. Consequently, the $\Neff$ formulated here represents a kernel-based effective sample size, not the definitive statistical degrees of freedom for the fully interacting flock. It provides an empirical comparison coordinate, not a parameter-free prediction of the interacting-flock transition.

\subsection{Circular statistics at high angular noise}
\label{app:circular}

The small-angle expansion above is not quantitatively controlled in the regime of the largest order enhancement: at $\eta=0.8$, the imposed angular error spans $\pm0.8\pi=\pm144^\circ$. A complementary calculation can be made without linearizing the noise angle. For one noisy unit phasor $z=e^{i(\theta_0+\xi)}$ with a fixed reference direction $\theta_0$ and $\xi\sim U(-\eta\pi,\eta\pi)$, its mean is exactly
\begin{equation}
\avg{z}=e^{i\theta_0}\lambda(\eta),\qquad
\lambda(\eta)=\frac{\sin(\pi\eta)}{\pi\eta},
\label{eq:lambda}
\end{equation}
where $\lambda(0)=1$ by continuity. If observations in the memory are \emph{independent} and share the same underlying direction, the normalized weighted mean $Z=\sum_m w_mz_m$ obeys
\begin{equation}
\avg{Z}=e^{i\theta_0}\lambda,\quad
\avg{|Z-\avg{Z}|^2}=\frac{1-\lambda^2}{\Neff}.
\label{eq:circular_variance}
\end{equation}
Thus the magnitude of the mean relative to the root-mean-square two-component fluctuations is
\begin{equation}
\mathcal S=\frac{\lambda\sqrt{\Neff}}{\sqrt{1-\lambda^2}}.
\label{eq:circular_snr}
\end{equation}
A still simpler exploratory coordinate, $\lambda\sqrt{\Neff}$, is often useful when the denominator varies weakly. Evaluated \emph{at the already reported finite-system midpoints}, it gives approximately $0.87$ for FBM $K=0.5$, $0.89$ for RLM $\alpha=0.4$, $0.91$ for FBM $K=0.05$ and RLM $\alpha=0.1$, and $0.85$ for FBM $K=0$; the near-memoryless midpoint gives about $0.70$. This clustering is suggestive, not a tested collapse or a fitted scaling law. In particular, Eq.~(\ref{eq:circular_variance}) is exact only for the stated independent-phasor problem; in the FBM flock the local direction evolves and successive observations are correlated, while the RLM averages vectors of nonconstant magnitude. The equation therefore does not predict $\eta_{50}$ from first principles in either interacting system. At $\eta=1$ the mean of a single noisy phasor vanishes, but finite-size and finite-time effects need not disappear exactly.

\subsection{Effective memory of the finite exponential buffer}

For the FBM model, $a_m=q^m$ where $q=e^{-K}$ and $m=0,\ldots,M-1$. The constituent geometric sums are
\begin{equation}
S_1=\sum_{m=0}^{M-1}q^m=\frac{1-q^M}{1-q},
\end{equation}
and
\begin{equation}
S_2=\sum_{m=0}^{M-1}q^{2m}=\frac{1-q^{2M}}{1-q^2}.
\end{equation}
Substituting into Eq.~(\ref{eq:neff_unnorm}) yields
\begin{align}
\Neff^{\rm buf}
&=\frac{S_1^2}{S_2}\\
&=\frac{1+q}{1-q}\frac{1-q^M}{1+q^M}.
\label{eq:neffbuf_closed}
\end{align}
The limiting behaviors trace $\Neff^{\rm buf}\to1$ as $K\to\infty$, and $\Neff^{\rm buf}\to M$ as $K\to0$. For a buffer of $M=21$, the maximum theoretical effective memory is precisely 21. For an operational decay of $K=0.05$, Eq.~(\ref{eq:neffbuf_closed}) outputs $\Neff^{\rm buf}\simeq19.266$.

\subsection{Effective memory of the recursive learning rule}

Write $\bm H_n\equiv\bm H(n\Delta t)$ and $\widetilde{\bm V}^{\mathrm{loc}}_n\equiv\widetilde{\bm V}^{\mathrm{loc}}(n\Delta t)$. After $n$ recursive updates starting from $\bm H_0$, Eq.~(\ref{eq:ema}) gives the exact finite-history identity
\begin{equation}
\begin{split}
\bm H_n={}&(1-\alpha)^n\bm H_0\\
&+\alpha\sum_{m=0}^{n-1}(1-\alpha)^m
\widetilde{\bm V}^{\mathrm{loc}}_{n-1-m}.
\end{split}
\label{eq:rlm_iterated}
\end{equation}
For $0<\alpha\leq1$ and after the initialization term has decayed, the corresponding stationary temporal kernel is $w_m=\alpha(1-\alpha)^m$, normalized over $m\geq0$. The sum of its squares evaluates to
\begin{equation}
\sum_{m=0}^{\infty}w_m^2
=\frac{\alpha}{2-\alpha},
\end{equation}
producing an effective memory of
\begin{equation}
\boxed{\Neff^{\rm RLM}=\frac{2-\alpha}{\alpha}}.
\end{equation}
Thus, an instantaneous update ($\alpha=1$) gives $\Neff=1$, while a highly persistent state ($\alpha=0.1$) yields $\Neff=19$.

\subsection{Matching the two kernels}

To algorithmically match a given FBM value $\Neff^{\rm buf}$, the required RLM persistence parameter is derived as
\begin{equation}
\boxed{\alpha_{\rm match}=\frac{2}{\Neff^{\rm buf}+1}}.
\end{equation}

Three physical caveats constrain the interpretation of this mathematical mapping. First, the derivation inherently linearizes the circular mean, rendering it most accurate for highly concentrated heading distributions (low effective noise). Second, it explicitly neglects temporal covariance, whereas dynamically interacting Vicsek flocks naturally generate correlated local spatial directions and temporally correlated effective errors. Third, the two memory rules aggregate entirely different microscopic properties: the FBM explicitly pools stored angles using fixed exponential weights, whereas the RLM recursively sums unnormalized local-alignment vectors whose physical magnitudes scale with dynamic local particle configurations. 

As a result, an identical $\Neff$ does not guarantee strictly identical flocking dynamics. Its utility here is empirical and comparative: the two architectures show similar steady-state polarization and operational midpoints at approximately matched nominal depth, without establishing equality of their transient dynamics.

\bibliography{references} 

\end{document}